\documentclass[hidelinks]{article}

\usepackage{PRIMEarxiv}

\usepackage[english]{babel}
\usepackage[centertags]{amsmath}
\usepackage{amssymb}
\usepackage{newlfont}

\newtheorem{theorem}{Theorem}[section]

\newtheorem{lemma}{Lemma}[section] 
\newtheorem{corollary}{Corollary}[section]
\newtheorem{definition}{Definition}[section]

\title{\large\bf Decision algorithms for fragments of real analysis.\ II. 
A theory of \\ differentiable functions with convexity and concavity
predicates\thanks{Work partially supported 
by MIUR project \emph{``Large-scale development 
of certified mathematical proofs''} n. 2006012773.}}

\author{Domenico Cantone$^{[0000-0002-1306-1166]}$ \\
  Dept.\ of Mathematics and Computer Science \\
  University of Catania \\
  I-95125 Catania, Italy\\
  \texttt{domenico.cantone@unict.it} \\
   \And
  Gianluca Cincotti$^{[0000-0001-8460-1708]}$\\
  Dept.\ of Mathematics and Computer Science \\
  University of Catania \\
  I-95125 Catania, Italy\\
  \texttt{cincotti@dmi.unict.it}
}

\newtheorem{myExample}{Example}

\newcommand{\defAs}
   {\mbox{$\:= \! \! \raisebox{-0.5 ex}[0 ex][0 ex]{\tiny Def}\:$}}

\newcommand{\equivAs}
   {\:\equiv \! \! \raisebox{-0.5 ex}[0 ex][0 ex]{\tiny Def}\:}

\newcommand{\comment}[1]{}

\newcommand{\Pfun}[5]{\mbox{$(#1  #2 #3)_{[#4,#5]}$}}
\newcommand{\Pder}[5]{\mbox{$(D[#1] #2 #3)_{[#4,#5]}$}}
\newcommand{\Pup}[3]{\mbox{Up$(#1)_{[#2,#3]}$}}
\newcommand{\Pdw}[3]{\mbox{Down$(#1)_{[#2,#3]}$}}
\newcommand{\Psup}[3]{\mbox{Strict\_Up$(#1)_{[#2,#3]}$}}
\newcommand{\Psdw}[3]{\mbox{Strict\_Down$(#1)_{[#2,#3]}$}}
\newcommand{\Pcx}[3]{\mbox{Convex$(#1)_{[#2,#3]}$}}
\newcommand{\Pcv}[3]{\mbox{Concave$(#1)_{[#2,#3]}$}}
\newcommand{\Pscx}[3]{\mbox{Strict\_Convex$(#1)_{[#2,#3]}$}}
\newcommand{\Pscv}[3]{\mbox{Strict\_Concave$(#1)_{[#2,#3]}$}}

\newcommand{\mif}{-\infty}
\newcommand{\pif}{+\infty}
\newcommand{\zm}{$z_1 = -\infty $ }
\newcommand{\zp}{$z_2 = +\infty $ }

\newcommand{\ind}[1]{\mbox{{\sl ind}$(#1)$}}
\newcommand{\infl}[2]{\in \{ \ind{#1}, \ld, \ind{#2} \}}
\newcommand{\influ}[2]{\in \{ \ind{#1}, \ld, \ind{#2}-1\}}

\newcommand{\uq}{{\textstyle{\frac{1}{4}}}}

\newcommand{\func}[1]{\mbox{\sl #1}}

\newcommand{\thh}{\theta}
\newcommand{\aels}{$[\al]$--elastics }
\newcommand{\cx}{${\cal C}(x) $ }
\newcommand{\um}{{\textstyle{\frac{1}{2}}}}

\newcommand{\rdf}{$\mathrm{RDF}$}
\newcommand{\rdford}{$\mathrm{RDF}_{ord}$}
\newcommand{\an}{\;\; \wedge \;\;}
\newcommand{\al}{\alpha}
\newcommand{\ga}{\gamma}
\newcommand{\ep}{\epsilon}
\newcommand{\gao}{\overline{\ga}}
\newcommand{\xx}{\xi}
\newcommand{\f}{\varphi}
\newcommand{\Ga}{\Gamma}

\newcommand{\y}{\overline{y}}
\newcommand{\tto}{\overline{t}}
\newcommand{\lea}{\leadsto}

\newcommand{\sra}{\rightarrow}
\newcommand{\ld}{\ldots}
\newcommand{\Rl}{{\mathbb R}}
\newcommand{\Rlp}{\Rl^{+}}
\newcommand{\Rlz}{\Rl_0^{+}}
\newcommand{\id}{\begin{proof}}
\newcommand{\fd}{\qed\end{proof}}
\newcommand{\op}{ \mbox{$ \bowtie $ } }
\newcommand{\spz}{\vspace{7 mm}}
\newcommand{\ev}[1]{{\it #1}}
\newcommand{\ed}[1]{{\emph{#1}}}
\newcommand{\ov}[1]{{\overline #1}}

\newcommand{\inn}[1]{\in \{1,2, \ld, #1 \}}

\newcommand{\sdef}[3]{ single-$[#1,#2,#3]$-defined }
\newcommand{\spdef}[3]{ special-$[#1,#2,#3]$-defined }
\newcommand{\ddef}[3]{ double-$[#1,#2,#3]$-defined }

\newcommand{\limplies}{\longrightarrow}

\newcommand{\Mm}[1]{\pmb{#1}}
\newcommand{\qed}{{\hfill\ensuremath{\Mm{\dashv}}}}

\begin{document}

\maketitle


\begin{abstract}
	We address the decision problem for a fragment of
	real analysis involving differentiable functions with
	continuous first derivatives.  The proposed theory, besides the
	operators of Tarski's theory of reals, includes predicates for
	comparisons, monotonicity, convexity, and derivative of
	functions over bounded closed intervals or unbounded
	intervals.

	Our decision algorithm is obtained by showing that satisfiable
	formulae of our theory admit canonical models in which
	functional variables are interpreted as piecewise exponential
	functions.  These can be implicitly described within the
	decidable Tarski's theory of reals.
	
	Our satisfiability test generalizes previous decidability
	results not involving derivative operators.

\smallskip

\noindent{\bf Key words:}\ Automated theorem proving, Decision procedures, 
	Elementary real analysis, Functions of a real variable.\\~
\noindent{\bf MS Classification 2020:}\quad {\small \textbf{03B25}, 26A99}.

\end{abstract}


\section{Introduction}
%
Verification of floating point hardware and hybrid systems has given
an important impulse to the formalization in computerized environments
of the theory of reals and some of its extensions.  In this
connection, among others, we cite the work on theorem proving with the
real numbers using a version of the HOL theorem prover
\cite{Harrison98}, the mechanization of real analysis in Isabelle/HOL
\cite{Fleuriot00}, in PVS \cite{Gottliebsen01}, and in the
interactive proof system IMPS \cite{GT93}, the ongoing efforts with
the Mizar system \cite{Bonarska90,Muzalewski93}, the attempt to
formalize Cauchy's integral theorem and integral formula in the
EtnaNova proof-verifier
\cite{SCO11,CanOmoSchUrs03,Omodeo-Schwartz-02}, 
and so on.

To keep within reasonable limits the amount of details that a user
must provide to a verification system in proof sessions, it is
necessary that the verifier has a rich endowment of decision
procedures, capable to formalize ``obvious'' deduction steps.  Thus, a
proof verifier for real analysis should include in its inferential
kernel a decision procedure for Tarski's elementary theory of reals
\cite{Tarski} as well as efficient decision tests for more
specialized subtheories such as the existential theory of reals
\cite{HeintzRoySolerno93}, the theory of bounded and stable quantified
constraints \cite{Ratschan06}, and other even more specific classes of
constraints.

In some situations, one may also need to reason about real functions,
represented in the language as interpreted or uninterpreted function
symbols.\footnote{Interpreted function symbols have a predefined
interpretation (e.g., the exponential and the sinus functions, $e^{x}$
and $\sin x$, respectively), whereas uninterpreted function symbols
have no predefined meaning attached to them and therefore they can be
interpreted freely (e.g. the ``generic'' function symbols $f$ and
$g$).}
However, 
one must be aware that the existential theory of reals extended with
the interpreted symbols $\log 2$, $\pi$, $e^{x}$, and $\sin x$ is
undecidable \cite{Richardson68}.  On the other hand, it has been shown
in \cite{MacintyreWilkie96} that the first-order theory of the real
numbers extended with the exponential function $e^{x}$ is decidable,
provided that Schanuel's conjecture in transcendental number theory
holds \cite[Chapter 3, pp.\ 145-176]{Chudnovsky84}.

The existential theory of reals has been extended in \cite{CFOS} with
uninterpreted continuous function symbols, function sum, function
point evaluation, and with predicates expressing comparison,
monotonicity (strict and non-strict) and non-strict convexity of
functions.  Such decidability result has been further extended in
\cite{CCG}, where a decision algorithm for the sublanguage RMCF$^+$
(augmented theory of Reals with Monotone and Convex Functions) of
elementary analysis has been given.  The theory RMCF$^+$ consists of
the propositional combination of predicates related to certain
properties of continuous functions which must hold in bounded or
unbounded intervals.  More precisely, the arguments of RMCF$^+$'s
predicates are numerical and functional terms.  Numerical variables
are interpreted by real numbers, whereas functional variables are
interpreted by everywhere defined continuous real functions of one
real variable.  Furthermore, numerical terms are obtained by composing
numerical variables through the basic arithmetic operations, whereas
functional terms are obtained by composing functional variables
through the addition (or subtraction) operator over functions.

\smallskip

In this paper, we consider a new theory, \rdf\ (theory of Reals and
Differentiable Functions), which in part extends the RMCF$^+$ theory
as it allows continuous and differentiable (with continuous
derivative) real functions as interpretation of functional variables.
The \rdf\ theory contains all RMCF$^+$'s predicates plus other
predicates concerning first order derivative of functions.

The arguments of \rdf's predicates are numerical terms and functional
variables (this is the only restriction with respect to the theory
RMCF$^+$, since addititive functional terms are not currently allowed
in \rdf).

We will show that the \rdf\ theory is decidable by generalizing the
proof techniques used in \cite{CCG,CFOS}.  In particular, the decision
algorithm is obtained by exploiting the fact that any satisfiable
\rdf-formula admits a model in which functional variables are
interpreted by parametric piecewise exponential functions. Since such 
functions can be implicitly described by existential formulae of 
Tarski's theory of reals, the decidability of \rdf\ follows.

 \spz

The paper is organized as follows.  In Section~\ref{sez_teor} we
present the syntax and semantics of \rdf-formulae.  Then in
Section~\ref{sez_norm} we review the normalization process for
unquantified formulae of first order theory with equality and
introduce the notion of ordered \rdf-formulae.  Section~\ref{sez_algo}
reports a decision algorithm for \rdf\ theory and a sketch proof of
its correctness.  
Final remarks and open problems are given in the last section.

\medskip

\noindent{\bf Historical note:}\ \emph{This paper was presented at the CILC'07
(Convegno Italiano di Logica Computazionale) held in Messina (21--22 June 2007);
it underwent anonymous reviews---the conference proceedings have not been published.}

\section{The \rdf\ theory} \label{sez_teor}

In this section we introduce the language
of the theory \rdf\ (Reals with Differentiable
Functions) and  give its intended semantics.

Two different kinds of variables can occur in \rdf.  \ev{Numerical
variables}, denoted with $x,y,\ld$, are used to represent real
numbers; whereas, \ev{functional variables}, denoted with $f,g,\ld$,
are used to represent continuous and differentiable (with continuous
derivative) real functions.  \rdf-formulae can also involve the
following constant symbols:
\begin{itemize}
	\item 0,1, which are interpreted as the real
	numbers 0,1, respectively;
	
	\item {\bf 0,1}, which are interpreted as the null function and
	the 1-constant function, respectively.
\end{itemize}
The language \rdf\ includes also two distinguished symbols: $\mif, \pif$;
they cannot occur everywhere in \rdf\ formulae but only as
``range defining'' parameters as it will be clear from the following
definitions.
\begin{definition}
	\ed{Numerical terms} are recursively defined by:
	\begin{enumerate}
  \item every numerical variable $x,y,\ld$ or constant 0,1 is a
  numerical term;
  
  \item if $t_1,t_2$ are numerical terms, then $(t_1+t_2), (t_1-t_2),
  (t_1*t_2)$, and $(t_1/t_2)$ are numerical terms;
  
  \item if $t$ is a numerical term and $f$ is a functional variable,
  then $f(t)$ and $D[f](t)$ are numerical terms;
  
  \item an expression is a numerical term only if it can be shown to
  be so on the basis of 1,2,3 above.
	\end{enumerate}
\end{definition}

In the following we will use the term ``functional variable''
to refer either	to a functional variable  or to a functional constant.

\begin{definition}
	An  \ed{extended numerical variable} (resp. term) is a numerical
	variable (resp. term) or one of the symbols  $\mif$ and $\pif$.
\end{definition}

\begin{definition} \label{def_atom}
	An \rdf-\ed{atom} is an expression having one of the following
	forms:
	\begin{eqnarray*}
		t_1 = t_2 &,& \; t_1 > t_2, \\
		\Pfun{f}{=}{g}{s_1}{s_2}	&,& \;  \Pfun{f}{>}{g}{t_1}{t_2} , \\
		\Pder{f}{=}{t}{s_1}{s_2}	&,& \;  \Pder{f}{>}{t}{s_1}{s_2} , \; \Pder{f}{\geq}{t}{s_1}{s_2} , \\
        \Pder{f}{<}{t}{s_1}{s_2} &,& \; \Pder{f}{\leq}{t}{s_1}{s_2} , \\
		\Pup{f}{s_1}{s_2}  &,& \; \Psup{f}{s_1}{s_2}, \\
		\Pdw{f}{s_1}{s_2}  &,& \; \Psdw{f}{s_1}{s_2}, \\
		\Pcx{f}{s_1}{s_2}  &,& \; \Pscx{f}{s_1}{s_2}, \\
		\Pcv{f}{s_1}{s_2}  &,& \; \Pscv{f}{s_1}{s_2},
	\end{eqnarray*}
	where  $t,t_1,t_2$ are numerical terms, $f, g$ are
	functional variables, and $s_1, s_2$ are extended numerical terms
	such that $s_1 \neq \pif$ and $s_2 \neq \mif$.
\end{definition}

\begin{definition}
	An \rdf-\ed{formula} is any propositional combination of \rdf-atoms.
\end{definition}

The semantics of \rdf-formulae is defined as follows.

\begin{definition}
A \ed{real model} for an \rdf-formula is an interpretation $M$ such
that:
\begin{enumerate}
	\item For any numerical variable $x$, the value $Mx$ is a real
	number.
			
	\item For any functional variable $f$, $(Mf)$ is an everywhere
	defined differentiable real function of one real variable with
	continuous derivative.
			
	\item For any composite numerical term $t_1 \otimes t_2$,
	$M(t_1 \otimes t_2)$ is the real number $Mt_1 \otimes Mt_2$,
	where $\otimes \in \{+,-,*,/\}$.
			
	\item For any numerical term $f(t)$, $M(f(t))$ is the real
	number $(Mf)(Mt)$.
	
	\item For any numerical term $D[f](t)$,
	$M(D[f](t))$ is the real number $D[(Mf)](Mt)$.
			
	\item Let $t,t_1,t_2$ be numerical terms, let $f,g$ be
	functional variables and assume that $Mt, Mt_1, Mt_2, (Mf),
	(Mg)$ are respectively their interpretations by $M$.  Let
	$s_1,s_2$ be extended numerical terms and let $Ms_1, Ms_2$ be
	their associated interpretations, where $Ms_i, i=1,2$ is a
	real number as above if $s_i$ is a numerical term; otherwise
	$Ms_i$ is the symbol $\mif$ (resp.  $\pif$) if $s_i=\mif$
	(resp.  $\pif$).

	\rdf-atoms are interpreted by a given real model $M$ according
	to the following rules:
	\begin{enumerate}
		\item $t_1=t_2$ (resp.  $t_1>t_2$) is true if and only
		if $Mt_1=Mt_2$ (resp.  $Mt_1 > Mt_2$);
				
		\item \Pfun{f}{=}{g}{s_1}{s_2} is true\footnote{With
		some abuse of notation, bounded and unbounded
		intervals are represented with the same formalism
		$[a,b]$.  Furthermore, we assume $\mif < \pif$.} if
		and only if $Ms_1>Ms_2$, or $Ms_1 \leq Ms_2$ and
		$(Mf)(x) = (Mg)(x)$ for every $x \in [Ms_1,Ms_2]$;
				
		\item \Pfun{f}{>}{g}{t_1}{t_2} is true if and only if
		$Mt_1>Mt_2$, or $Mt_1 \leq Mt_2$ and $(Mf)(x) >
		(Mg)(x)$ for every $x \in [Mt_1,Mt_2]$;
				
		\item \Pder{f}{\op}{t}{s_1}{s_2}, with $\op \in \{=,
		>, \geq, <, \leq\}$, is true if and only if
		$Ms_1>Ms_2$, or $Ms_1 \leq Ms_2$ and $D[(Mf)](x) \op
		Mt$ for every $x \in [Ms_1,Ms_2]$;
				
		\item \Pup{f}{s_1}{s_2} (resp.  Strict\_Up) is true if
		and only if $Ms_1 \geq Ms_2$, or $Ms_1 < Ms_2$ and the
		function $(Mf)$ is monotone nondecreasing (resp.
		strictly increasing) in the interval $[Ms_1,Ms_2]$;
				
		\item \Pcx{f}{s_1}{s_2} (resp.  Strict\_Convex) is
		true if and only if $Ms_1 \geq Ms_2$, or $Ms_1 < Ms_2$
		and the function $(Mf)$ is convex (resp.  strictly
		convex) in the interval $[Ms_1,Ms_2]$;
				
		\item the truth value of \Pdw{f}{s_1}{s_2} (resp.
		Strict\_Down) and \linebreak \Pcv{f}{s_1}{s_2} (resp.
		Strict\_Concave) are defined in a manner completely
		analogous to the above definitions (e) and (f).
	\end{enumerate}
\end{enumerate}
\end{definition}

One can not expect that any deep theorem of real analysis can be
directly expressed by an \rdf-formula, and therefore automatically
verified.  Indeed, our decidability result is to be regarded as just
one more step towards the mechanization of the ``obvious'', which is
basic for the realization of powerful interactive proof verifiers in
which the user assumes control only for the more challenging deduction
steps (such as, for instance, the instantiation of quantified
variables), otherwise leaving the burden of the verification of small
details to the system.

\medskip

We give next a few examples of statements which could be verified
automatically by our proposed decision test for \rdf.

\begin{myExample}
\label{ex1}
\emph{Let $f$ be a real differential function in a closed interval $[a,b]$
with continuous derivative such that $f(a) = f(b)$, $f'(a) \neq 0$,
and $f'(b) \neq 0$.  Then there exists some $a < c < b$ such that
$f'(c)=0$.  (This is a weakened version of Rolle's theorem.)
}

\bigskip

A possible formalization of the above statement is given by the
universal closure of the formula:
%
\begin{multline*}
  \left(a < b \land f(a)=f(b) \land D[f](a) \neq 0 \land D[f](b) \neq 0
  \right) 
  \limplies (\exists c)(a < c < b \land D[f](a) = 0)\,,
\end{multline*}
%
whose theoremhood can be tested by showing that the following \rdf-formula
is unsatisfiable:
\begin{multline*}
  \left(a < b \land f(a)=f(b) \land D[f](a) \neq 0 \land D[f](b) \neq 0
  \right) 
  \land\left(\Pder{f}{>}{0}{a}{b} \lor \Pder{f}{>}{0}{a}{b}\right)\,.
\end{multline*}
\qed
\end{myExample}

\begin{myExample}
\label{ex4} 
\emph{Let $f$ be a real differential function in a closed interval
$[a,b]$ with constant derivative in $[a,b]$.  Then the function $f$ is
linear in $[a,b]$.  }

\bigskip

A possible formalization of the above statement is given by the
universal closure of the formula:
\begin{equation*}
\Pder{f}{=}{t}{a}{b}  \limplies 
   \bigg(\Pcx{f}{a}{b} \land
   \Pcv{f}{a}{b} \bigg)\,,
\end{equation*}
whose theoremhood can be tested by showing that the following \rdf-formula
is unsatisfiable:
\begin{equation*}
\Pder{f}{=}{t}{a}{b}  \land  
   \left(\lnot \Pcx{f}{a}{b} \lor
   \lnot \Pcv{f}{a}{b} \right)\,.
\end{equation*}
\qed
\end{myExample}

In the rest of the paper we will present a satisfiability test for
\rdf.


\section{The normalization process}  \label{sez_norm}

Our decidability test makes use of the following general 
normalization process (cf.~\cite{CFOS}).
Let $T$ be an unquantified first order theory, with equality
=, variables $x_1,x_2,\ld$, function symbols
$f_1,f_2,\ld$ and predicate symbols $P_1,P_2, \ld$.
\begin{definition}
	A formula $\f$ of $T$ is in \ed{normal form} if:
	\begin{enumerate}
		\item every term occurring in $\f$ is a variable or
		has the form $f(x_1,x_2,\ld,x_n)$, where
		$x_1,x_2,\ld,x_n$ are variables and $f$ is a function
		symbol;
 
		\item every atom in $\f$ is in the form $x=t$ where
		$x$ is a variable and $t$ is a term, or in the form
		$P(x_1,x_2,\ld,x_n)$, where $x_1,x_2,\ld,x_n$ are
		variables and $P$ is a predicate symbol.
	\end{enumerate}
\end{definition}

\begin{lemma}
	There is an effective procedure to transform any formula $\f$
	in $T$ into an equisatisfiable formula $\psi$ in normal form.
\end{lemma}
\id
	See \cite[Lemma 2.2]{CFOS}.
\fd

\begin{definition}
	A normal form formula $\f$ of $T$ is in
	\ed{standard normal form} if it is a conjunction of literals
	of the kinds:
	\[
		x=y , \quad x=f(x_1,\ld,x_n), \quad x \neq y
	\]
	\[
  P(x_1,x_2,\ld,x_n) , \quad		\neg P(x_1,x_2,\ld,x_n) .
	\]
	where $x,y,x_1,x_2,\ld,x_n$ are variables, $f$ is a function
	symbol, and $P$ is a predicate symbol.
\end{definition}

Let $S$ be the class of all formulae of $T$ in standard normal form. 
Then we have:
\begin{lemma} \label{lem_TS}
	$T$ is decidable if and only if $S$ is decidable.
\end{lemma}
\id
	See \cite[Lemma 2.4]{CFOS}.
\fd

Next, we describe the standard normal form 
for the \rdf\ theory.

First of all, we observe that the following equivalences hold:
\begin{eqnarray*}
	t_1=t_2-t_3	\; & \equiv & \;	t_2=t_1+t_3 , \\
	t_1=t_2/t_3	\; & \equiv & \;	(t_3 \neq 0) \an (t_2=t_1*t_3), \\
	t_1 \neq t_2	\; & \equiv & \;	(t_2>t_1) \vee (t_1>t_2),		\\
	t_1 \not> t_2 \; & \equiv & \;	(t_1=t_2) \vee (t_2>t_1),		\\
	t_1>t_2			\; & \equiv & \;	(t_1=t_2+v) \an (v>0),
\end{eqnarray*}
where $t_1,t_2,t_3$ stand for numerical terms and $v$ stands for a new
numerical variable.  Moreover, by recalling that function symbols are
modeled by differentiable functions (with continuous derivative), the
following equivalences hold:
 \begin{eqnarray*}
 \Pup{f}{s_1}{s_2} &\equiv&  \; \Pder{f}{\geq}{0}{s_1}{s_2},  \\
 \Pdw{f}{s_1}{s_2} &\equiv&  \; \Pder{f}{\leq}{0}{s_1}{s_2} ,
 \end{eqnarray*}
where $s_1,s_2$ are extended numerical terms.\footnote{Observe that
strict-monotonicity predicates can not be treated in the same way.}

By applying the elementary normalization process hinted to above, we
can consider, w.l.o.g., only formulae in \ev{standard normal form}
whose literals are of the following kinds:
\begin{eqnarray*}
  x=y+w ,\quad  x=y*w , \quad &x>0& , \quad y=f(x) , \quad y=D[f](x),\\
  \Pfun{f}{=}{g}{z_1}{z_2}, \quad \Pfun{f}{\neq}{g}{z_1}{z_2},&&\\
  \Pfun{f}{>}{g}{w_1}{w_2}, \quad \Pfun{f}{\not>}{g}{w_1}{w_2},&&\\
  \Pder{f}{\op}{y}{z_1}{z_2}, \quad \Pder{f}{\not \op}{y}{z_1}{z_2},&\quad&
	\mbox{ where } \;  \op \in \{=, >, \geq, <, \leq\},\\
  \Psup{f}{z_1}{z_2} , \quad  \neg \Psup{f}{z_1}{z_2} ,
 &\quad& \mbox{(resp. Strict\_Down)},\\
  \Pcx{f}{z_1}{z_2} ,	\quad \neg \Pcx{f}{z_1}{z_2} ,
 &\quad& \mbox{(resp. Strict\_Convex)},\\
  \Pcv{f}{z_1}{z_2} ,	\quad \neg \Pcv{f}{z_1}{z_2} ,
 &\quad& \mbox{(resp. Strict\_Concave)},
\end{eqnarray*}
where $x,y,w,w_1,w_2$ are numerical variables, $z_1,z_2$ are extended
numerical variables,\footnote{By Definition~\ref{def_atom}, we have
$z_1 \neq \pif$ and $z_2 \neq \mif$.} and $f,g$ are functional
variables.

\spz

The decision algorithm to be presented in Section~\ref{sez_algo}
requires one more preparatory step.

Let $\f$ be a formula of the \rdf\ theory.
\begin{definition}
	A \ed{domain variable} for a formula $\f$ is a numerical
	variable such that it is either the argument of some
	functional variable (e.g., $x$ in $f(x)$ or in $D[f](x)$) or
	it is the argument of the range defining parameters of some
	interval mentioned in $\f$ (e.g., $x$ and $y$ in
	$\Pcx{f}{x}{y}$).
\end{definition}

We consider a \ev{strict linear ordering} of the
domain variables:
\begin{definition}		\label {def_ord}
	Let $D = \{ x_1,x_2,\ld,x_n \}$ be the set of domain variables
	of $\f$.  A formula $\f$ is said to be \ed{ordered} if:
	\[
		\f \mbox{ is satisfiable} \quad \Longleftrightarrow \quad 
		\f \wedge \bigwedge_{i=1}^{n-1} (x_i < x_{i+1}) \mbox{ is satisfiable}.
	\]
\end{definition}

The family \rdford\ of all the ordered formulae in \rdf\
is a proper subset of \rdf.  It is straightforward to
prove that:

\begin{lemma} \label{lem_ord}
	\rdf\ is decidable if and only if \rdford\ is decidable. \qed
\end{lemma}

In the rest of the paper, formulae of \rdf\ will always be assumed to
be ordered and in standard normal form.


\section{The decision algorithm}  \label{sez_algo}

In this section we present a decision algorithm which solves the
satisfiability problem for \rdf-formulae.  The algorithm takes as
input an ordered \rdf-formula $\f$ in standard normal form and
reduces it, through a sequence of effective and satisfiability
preserving transformations $\f \lea \f_1 \lea \f_2 \lea \f_3$, into a
formula $\psi = \f_{3}$ of the existential Tarski's theory of reals,
i.e., an existentially quantified formula involving only real
variables, the arithmetic operators $+, *$, and the predicates $=, <$.

From the decidability of Tarski's theory of reals (see \cite{Tarski};
see also \cite{Collins}), the decidability of \rdf\ follows 
immediately.

\smallskip

In the following, $w_i$ denotes a numerical variable, whereas
$z_i$ denotes an extended numerical variable.

In view of the results mentioned in the previous section, without loss
of generality we can assume that our input formula $\f$ is in standard
normal form and ordered.  The sequence of transformations needed to go
from $\f$ to $\psi$ is given by the following:


\paragraph{\underline{Reduction algorithm for \rdf.}} \mbox { } \\
-- {\sf Input} : an ordered formula $\f$ of \rdf\ in standard
normal form.  \\
-- {\sf Output} : a formula $\f_3$ of the Tarski's theory of reals
which is equisatisfiable with $\f$.  \\
The algorithm involves the following three fundamental steps:
\begin{enumerate}
\item  $\f \lea \f_1$: {\em Negative clauses removal.} \\
	Given an ordered formula $\f$ of \rdf\ in standard normal
	form, we construct an equisatisfiable formula $\f_1$ involving
	only positive predicates.  The general idea applied in this
	step is to substitute every negative clause involving a
	functional symbol with an implicit existential assertion.

	For the sake of simplicity, in the following we use the
	relation $x \preceq y$ as a shorthand for 
	\[
	x \preceq y \equivAs 
	  \begin{cases}
	        x \leq y & \hbox{if $x$ and $y$ are both numerical 
		variables}\\
		\mathbf{true} & \hbox{if $x =  \mif$ or $y = \pif$}\\
		\mathbf{false} & \hbox{if $x =  \pif$ and $y = \mif$}.
	  \end{cases}
	\]

	\begin{enumerate}
	\item For any literal of type \Pfun{f}{\neq}{g}{z_1}{z_2} occurring in
	$\f$, introduce three new numerical variables $x,y_1,y_2$ and
	replace \Pfun{f}{\neq}{g}{z_1}{z_2} by the formula:
	 \[
	 \Ga \an y_1 \neq y_2 ,
	 \]
	 where
	 \[
	 \Ga \equiv (z_1 \preceq x \preceq z_2) \an y_1=f(x) \an y_2=g(x).
	 \]

	 \item Replace any literal of type  \Pfun{f}{\not >}{g}{w_1}{w_2} occurring
	 in $\f$  by the formula:
	 \[
	 \Ga \an y_1 \leq y_2,
	 \]
	 where
	 \[
	 \Ga \equiv (w_1 \leq x \leq w_2) \an y_1=f(x) \an y_2=g(x).
	 \]
	 and $x,y_1,y_2$ are new numerical variables.

	 \item Replace any literal  of type \Pfun{D[f]}{\neq}{y}{z_1}{z_2}
	 (resp. $\not>, \not\geq, \not<, \not\leq$) occurring
	 in $\f$ by the formula:
	 \[
	 \Ga \an y_1 \neq y \qquad (\mbox{resp. } \leq, <, \geq, >) ,
	 \]
	 where
	 \[
	 \Ga \equiv (z_1 \preceq x \preceq z_2) \an y_1=D[f](x) ,
	 \]
	 and $x,y_1$ are new numerical variables.

	 \item	Replace any literal of type $\neg \Pup{f}{z_1}{z_2}$  (resp.
	 $\neg \Psup{f}{z_1}{z_2}$) occurring in $\f$ by the
	 formula:
	 \[
	   \Ga \an y_1 > y_2 \qquad 
	   (\mbox{resp.  } \Ga \an y_1 \geq y_2),
	 \]
	 where
	 \[
	 \Ga \equiv  (z_1 \preceq x_1 < x_2	\preceq z_2)
	 \an \bigwedge_{i=1}^{2} y_i = f(x_i),
	 \]
	 and $x_1,x_2$, $y_1,y_2$ are new
	 numerical variables.

	 The case of literals of the form $\neg \Pdw{f}{z_1}{z_2}$
	 (resp.  $\neg \Psdw{f}{z_1}{z_2}$) can be handled similarly.

	\item  Replace any literal of type $\neg \Pcx{f}{z_1}{z_2}$	(resp.
	 $\neg \Pscx{f}{z_1}{z_2}$) occurring in $\f$ by the
	 formula:\footnote{The formula asserts
	  the existence of three points $x_1,x_2,x_3$, such that
	  $(x_2,f(x_2))$ lies above (resp. $(x_2,f(x_2))$ does not lie below)
	  the straight line joining the two points
	  $(x_1,f(x_1))$ and $(x_3,f(x_3))$.}
	 \[
					\Ga \an (y_2-y_1)(x_3-x_1) > (x_2-x_1)(y_3-y_1)
	 \]
	 \[
					(\mbox{resp. } \Ga \an
	 (y_2-y_1)(x_3-x_1) \geq (x_2-x_1)(y_3-y_1)),
	 \]
	 where,
	 \[
					\Ga \equiv (z_1 \preceq  x_1 < x_2 < x_3 \preceq z_2)
						\an	\bigwedge_{i=1}^{3}	y_i = f(x_i)
	 \]
	 and $x_1,x_2,x_3,$ $y_1,y_2,y_3$ are new
	 numerical variables.

	 The case of of literals of the form $\neg \Pcv{f}{z_1}{z_2}$ (resp.
	 $\neg \Pscv{f}{z_1}{z_2}$) can be handled similarly.
	\end{enumerate}
	
    It is straightforward to prove that the formulae $\f$ and $\f_1$
    are equisatisfiable.
    
    \smallskip

    At this point, by normalizing $\f_1$ and in view of
    Lemma~\ref{lem_TS}, we can assume without loss of generality that
    $\f_1$ is in standard normal form, i.e., it is a conjunction of
    literals of the following types:
    \begin{eqnarray*}
	      x=y+w ,\quad  x=y*w , &\quad& x>0 , \quad y=f(x) , \quad y=D[f](x),\\
	      \Pfun{f}{=}{g}{z_1}{z_2}, &\quad&  \Pfun{f}{>}{g}{w_1}{w_2},\\
	      \Pder{f}{\op}{y}{z_1}{z_2}, &\quad&
	      \mbox{ where } \; \op \in \{=, >, \geq, <, \leq\},\\
			    \Psup{f}{z_1}{z_2}, &\quad& \Psdw{f}{z_1}{z_2},\\
			    \Pcx{f}{z_1}{z_2} , &\quad& \Pscx{f}{z_1}{z_2},\\
			    \Pcv{f}{z_1}{z_2} , &\quad& \Pscv{f}{z_1}{z_2}.
    \end{eqnarray*}
    Furthermore, in view of Lemma~\ref{lem_ord}, we may further assume
    that $\f_1$ is ordered with domain variables $v_1,v_2,\ld,v_r$.

\item  $\f_1 \lea \f_2$: {\em Explicit evaluation of functions
  over domain variables.}

 This step is in preparation of the elimination of functional clauses.

 For every domain variable $v_j$ and for every functional variable $f$
 occurring in $\f_1$, introduce two new numerical variables $y_j^f$,
 $t_j^f$ and add the literals $y_j^f=f(v_j)$ and $t_j^f=D[f](v_j)$ to
 $\f_1$.  Moreover, for each literal $x=f(v_j)$ already occurring in
 $\f_1$, add the literal $x=y_j^f$; likewise, for each literal
 $x=D[f](v_j)$ already occurring in $\f_1$, add the literal $x=t_j^f$.
 \\
 In this way, we obtain a new formula $\f_2$ which is
 clearly equisatisfiable to $\f_1$.

\item  $\f_2 \lea \f_3$: {\em Functional variables removal.}

	In this step all literals containing functional variables
	are eliminated.

	Let $V = \{v_1,v_2,\ld,v_r\}$ be the collection of the domain
	variables of $\f_2$ with their implicit ordering, and let the
	index function {\sl ind} : $V \cup \{\mif, \pif\} \longmapsto
	\{1,2, \ld , r\}$ be defined as follows:
	\[
	     \ind{x} \defAs
	      \begin{cases}
	     1  &  \mbox{if $x = - \infty$,} \\
	     l  &  \mbox{if $x = v_l$,~  for some $l \inn{r}$,} \\
	     r  &  \mbox{if $x = + \infty$.}
	      \end{cases}
	\]

	For each functional symbol $f$ occurring in $\f_2$, let us
	introduce the new numerical variables $\ga_0^f, \ga_r^f$ and
	proceed as follows:
	\begin{enumerate}
	\item For each literal of type \Pfun{f}{=}{g}{z_1}{z_2}
	occurring in $\f_2$, add the literals:
	 \[
					y_i^f = y_i^g,  \qquad \quad t_i^f = t_i^g,
	 \]
	 for $i \infl{z_1}{z_2}$;
	 moreover, if $z_1 = - \infty$, add the literal:
	 \[
					\ga_0^f = \ga_0^g\,;
	 \]
	 likewise, if $z_2 = + \infty$, add the literal:
	 \[
					\ga_r^f = \ga_r^g \,.
	 \]

	\item  For each literal of type \Pfun{f}{>}{g}{w_1}{w_2} occurring in $\f_2$,
	 add the literal:
	 \[
					y_i^f > y_i^g\,,
	 \]
	 for $i \infl{w_1}{w_2}$.

	\item  For each literal of type \Pder{f}{\op}{y}{z_1}{z_2}
	 occurring in $\f_2$, where $\op \in$ \mbox{$\{=,<,\leq,>,\geq\}$}, add
	 the formulae:
	 \[
					t_i^f \op y\,,
	 \]
	 \[
	 \frac{y_{j+1}^f-y_j^f}{v_{j+1}-v_j} \op y ,
	 \]
	 for $i, j \infl{z_1}{z_2}$, $j \neq \ind{z_2}$.
	 Additionally, if $\op \in \{\leq,\geq\}$ add also the
	 formulae:
	 \[
	 \left( \frac{y_{j+1}^f-y_j^f}{v_{j+1}-v_j} = y  \right)
	  \longrightarrow  (t_j^f=y \an t_{j+1}^f=y) ;
	 \]
	 moreover, if \zm, add the formula:
	 \[
					\ga_0^f \op y,
	 \]
	 and if \zp, add the formula:
	 \[
					\ga_r^f \op y.
	 \]

	  \item For each literal of type \Psup{f}{z_1}{z_2} (resp.
	  \Psdw{f}{z_1}{z_2}) occurring in $\f_2$, add the
	  formulae:
	 \[
					t_i^f \geq 0	\qquad  \mbox{(resp. } \leq),
	 \]
	 \[
	 y_{j+1}^f > y_j^f	\qquad \mbox{(resp. } <),
	 \]
	 for $i, j \infl{z_1}{z_2}$, $j \neq \ind{z_2}$.
	 Moreover, if \hbox{\zm,} add the formula:
	 \[
					\ga_0^f > 0  \qquad		\mbox{(resp. } <),
	 \]
	 and if \zp, add the formula:
	 \[
					\ga_r^f > 0	\qquad	\mbox{(resp. } <).
	 \]

	\item For each literal of type \Pcx{f}{z_1}{z_2} (resp.
	\Pcv{f}{z_1}{z_2}) occurring in $\f_2$, add the following
	formulae:\footnote{Observe that this group of formulae
	implicitly forces the relations $\frac{y_j^f -
	y_{j-1}^f}{v_j-v_{j-1}} \leq
	\frac{y_{j+1}^f-y_j^f}{v_{j+1}-v_j}$ for each $j \in
	\{\ind{z_1}+1, \ld , \ind{z_2}-1\}$.  Geometrically, the point
	of coordinates $(v_j,y_j^f)$ does not lie above (resp.  lies
	below) the straight line joining the two points $(v_{j-1},
	y_{j-1}^f)$ and $(v_{j+1}, y_{j+1}^f)$.}
	 \[
	 t_i^f \leq  \frac{y_{i+1}^f-y_i^f}{v_{i+1}-v_i}
	 \leq  t_{i+1}^f  \qquad
	  \mbox{(resp. } \geq),
	 \]
	 \[
		\left( \frac{y_{i+1}^f-y_i^f}{v_{i+1}-v_i} = t_i^f  \;\vee\;
	 \frac{y_{i+1}^f-y_i^f}{v_{i+1}-v_i} = t_{i+1}^f	\right)
	  \longrightarrow	(t_i^f= t_{i+1}^f)  ,
	 \]
	 for $i \influ{z_1}{z_2}$; moreover, if \zm, add the
	 formula:
	 \[
		\ga_0^f \leq t_1^f	\qquad  \mbox{(resp. } \geq),
	 \]
	 and if \zp,	add the formula:
	 \[
		\ga_r^f \geq t_r^f	\qquad  \mbox{(resp. } \leq).
	 \]

	\item  For each literal of type \Pscx{f}{z_1}{z_2} (resp.
	 \Pscv{f}{z_1}{z_2}) occurring in $\f_2$,
	 add the following formulae:
	 \[
	 t_i^f <  \frac{y_{i+1}^f-y_i^f}{v_{i+1}-v_i} < t_{i+1}^f \qquad
	  \mbox{(resp. } >),
	 \]
	 for $i \influ{z_1}{z_2}$;
	 moreover, if \zm, add the formula:
	 \[
		\ga_0^f <	t_1^f \qquad	\mbox{(resp. } >),
	 \]
	 and if \zp,	add the formula:
	 \[
		\ga_r^f >	t_r^f \qquad	\mbox{(resp. } <).
	 \]

	\item Drop all literals involving some functional variable.
	
	Let $\f_3$ be the resulting formula.
	\end{enumerate}
	
\end{enumerate}

Plainly, the formula $\f_3$ involves only literals
of the following types:
\[
    t_{1} \leq t_{2}\,, \qquad t_{1} < t_{2}\,, \qquad t_{1} = 
    t_{2}\,,
\]
where $t_{1}$ and $t_{2}$ are terms involving only real variables, the
real constants $0$ and $1$, and the arithmetic operators $+$ and $*$
(and their duals $-$ and $/$), so that the formula $\f_3$ belongs to
the decidable (existential) Tarski's theory of reals.
Thus, to show that our theory \rdf\ has a solvable satisfiability
problem, it is enough to show that even the latter transformation step
preserves satisfiability.  Such result, though not particularly deep,
requires a quite technical and lenghty proof, which will be provided 
in an extended version of the present paper.
 
 Therefore we can conclude with our main result:
 \begin{theorem}
     The theory \rdf\ has a decidable satisfiability problem. \qed
 \end{theorem}


\section{Conclusions}

In this paper we have shown that the satisfiability problem for the
fragment \rdf\ of the theory of reals with differential functions and
various predicates is solvable.  

The result has been obtained through a sequence of effective
reductions which transforms any \rdf-formula into an equisatisfiable
formula of the decidable Tarski's theory of reals.  Our decidability
result is based on the existence of canonical models of satisfiable
\rdf-formulae in which functional variables are interpreted by
parametric piecewise exponential functions.  

By further generalizing canonical models, we expect that more
elaborate constructs (such as, for instance, function addition and
derivation predicate over open intervals) can be allowed without
disrupting decidability.


\bibliographystyle{plain}

\begin{thebibliography}{RDF}




\bibitem{Bonarska90}
E.~Bonarska.  \emph{An Introduction to PC Mizar}, Fondation Philippe
le Hodey, Brussels, 1990.


\bibitem{CFO-book89}
D.~Cantone, A.~Ferro, and E.~G. Omodeo.
\newblock {\em Computable set theory}, volume no.6 Oxford Science Publications
  of {\em International Series of Monographs on Computer Science}.
\newblock Clarendon Press, 1989.

\bibitem{CCG}
	D. Cantone,  G. Cincotti, G. Gallo.
	\newblock  Decision algorithms for some fragments of real
					analysis: A theory of continuous functions with
					strict-convexity constructs.
	\newblock  {\it Journal of Symbolic Computation}, 
        Vol. 41(7), pp.\ 763--789, 2006.

\bibitem{CFOS}
	D. Cantone , A. Ferro, E. Omodeo, J.T. Schwartz.
	\newblock		Decision algorithms for some fragments of
  analysis and related	areas.
	\newblock {\it		Communications on Pure and Applied Mathematics},
  Vol.40, pp.\ 281-300, 1987.

\bibitem{CanOmoSchUrs03}
D.~Cantone, E.G.~Omodeo, J.T.~Schwartz, and P.~Ursino.
Notes from the Logbook of a Proof-Checker's Project.
In N.~Dershowitz, editor, \emph{Proc.\ of the International Symposium on Verification: Theory and Practice},
LNCS 2772, pp.\ 182--207, Springer-Verlag, 2003.

\bibitem{Chudnovsky84}
G.V. Chudnovsky.  
\emph{Contributions to the Theory of Transcendental Numbers}.
Amer.\  Math.\  Soc.\, Providence, RI, 1984.


\bibitem{Collins}
	G. Collins.
	\newblock Quantifier elimination for real closed fields
  by cylindrical algebraic decomposition.
	\newblock In {\it Second GI Conference on Automata Theory and Formal Languages},
  LNCS Vol. 33, Springer-Verlag, Berlin, 1975.

\bibitem{FOS}
	A. Ferro, E.G. Omodeo, J.T. Schwartz.
	\newblock  Decision procedures  for elementary sublanguages
  of set theory, I. Multi-level syllogistic and some extensions.
	\newblock {\it Communications on Pure and 
	Applied Mathematics},	Vol.\ 33, pp.\ 599--608, 1980.

\bibitem{Fleuriot00}
J.D. Fleuriot.  On the Mechanization of Real Analysis in Isabelle/HOL.
In \emph{Proc.\ of the 13th International Conference on Theorem Proving in
Higher Order}, LNCS 1869, pp.\ 145--161, Springer-Verlag, London, UK, 
2000.

\bibitem{Gottliebsen01}
   H. Gottliebsen. 
   \emph{Automated Theorem Proving for Mathematics: Real Analysis in PVS}. 
   PhD thesis, University of St.\ Andrews, 2001.

\bibitem{GT93}
J.D. Guttman and F.J. Thayer. 
IMPS: An Interactive Mathematical Proof System.
\emph{Journal of Automated Reasoning}, 11:213-248, 1993.

\bibitem{Harrison98}
J.~Harrison. \emph{Theorem Proving with the Real Numbers}, 
Springer-Verlag, 1998.

\bibitem{HeintzRoySolerno93}
J. Heintz, M.-F. Roy, and P. Solern\'o. %
On the theoretical and practical complexity of the existential theory
of reals. %
\emph{The Computer Journal}, Vol.  36 (5), pp.\  427-431, 1993.






\bibitem{MacintyreWilkie96}
A.J.~Macintyre and A.J.~Wilkie,
\newblock On the decidability of the real exponential field.
In P.~Odifreddi, editor, \emph{Kreiseliana},
pp.\ 441--467, A.K.~Peters, Wellesley, MA, 1996.

\bibitem{Muzalewski93}
M.~Muzalewski. \emph{An Outline of PC Mizar}, Fondation Philippe le Hodey, Brussels, 1993.


\bibitem{Omodeo-Schwartz-02}
E.G. Omodeo and J.T. Schwartz.
\newblock A `{t}heory' mechanism for a proof-verifier based on first-order set
  theory.
\newblock In A.~Kakas and F.~Sadri, editors, {\em Computational Logic: Logic
  Programming and Beyond}, LNCS 2408, pp.\ 214--230, Springer-Verlag, 2002.


\bibitem{Quine}
	W.V. Quine.
	\newblock  {\em Methods of logic}.
	\newblock  Henry Holt, New York,	1950.



\bibitem{Ratschan06}
S. Ratschan.  Efficient solving of quantified inequality constraints
over the real numbers.  To appear in \emph{ACM Transactions on
Computational Logic}. A pre-print is available at the \textsf{arXiv.org 
e-Print} archive at \texttt{http://arxiv.org/}.

\bibitem{Richardson68}
D. Richardson.  Some undecidable problems involving elementary
functions of a real variable.  \emph{Journal of Symbolic Logic}, Vol.\
33, pp.\  514--520, 1968.

\bibitem{SCO11}
Schwartz J.T., Cantone D., Omodeo E.G., 
{\rm Computational Logic and Set Theory -- \small Applying Formalized
  Logic to Analysis}.
Springer, 2011.
Foreword by Martin Davis.


\bibitem{TPTP}
   G. Sutcliffe and C.B. Suttner,
   \newblock The TPTP Problem Library: CNF Release v1.2.1,
   \newblock {\em Journal of Automated Reasoning}, 
   Vol. 21(2), pp. 177--203, 1998.
   \newblock (Available online at \texttt{http://www.tptp.org})
  
\bibitem{Tarski}
	A. Tarski.
	\newblock {\em A Decision method for elementary algebra
  and geometry} (2nd ed. rev.).
	\newblock University of California Press, Berkeley,	1951.
	 
 

\end{thebibliography}

\end{document}